\documentclass[aps,prl,twoside,twocolumn,floatfix,amsmath,showpacs,amssymb]{revtex4}

 \usepackage[latin1]{inputenc}
 \usepackage{amsmath}
 \usepackage{subfigure}
 \usepackage{placeins}
 \usepackage{floatflt}
 \usepackage{amsmath}
 \usepackage{amssymb}
 \usepackage{wrapfig}
 \usepackage[]{tocbibind}
 \usepackage[all]{xy}

 \usepackage{ifthen}
 \usepackage{times}

\makeatletter
\newcommand\figcaption{\def\@captype{figure}\caption}
\newcommand\tabcaption{\def\@captype{table}\caption}
\makeatother

\usepackage[pdftex]{graphicx}                                

\begin{document}

\DeclareGraphicsExtensions{.jpg,.pdf,.png,.mps,.eps,.ps}  

\title{ $\mathbf{\Lambda}$p femtoscopy in collisions of Ar+KCl at 1.76\boldmath{$A$}~GeV}

\author{
G.~Agakishiev$^{8}$, A.~Balanda$^{3}$, R.~Bassini$^{9}$,
D.~Belver$^{15}$, A.V.~Belyaev$^{6}$, A.~Blanco$^{2}$, M.~B\"{o}hmer$^{11}$,
J.\,L.~Boyard$^{13}$, P.~Braun-Munzinger$^{4}$, P.~Cabanelas$^{15}$
% ...
\\[5bp]
(HADES collaboration)
}

\affiliation{
\\ \vspace*{5bp} \mbox{$^{1}$Istituto Nazionale di Fisica Nucleare - Laboratori Nazionali del Sud,
95125~Catania, Italy}\\
\mbox{$^{2}$LIP-Laborat\'{o}rio de Instrumenta\c{c}\~{a}o e F\'{\i}sica
Experimental de Part\'{\i}culas , 3004-516~Coimbra, Portugal}\\
\mbox{$^{3}$Smoluchowski Institute of Physics, Jagiellonian University of Cracow,
30-059~Krak\'{o}w, Poland}\\
\mbox{$^{4}$GSI Helmholtzzentrum f\"{u}r Schwerionenforschung GmbH,
64291~Darmstadt, Germany}\\
\mbox{$^{5}$Institut f\"{u}r Strahlenphysik, Forschungszentrum Dresden-Rossendorf,
01314~Dresden, Germany}\\
\mbox{$^{6}$Joint Institute of Nuclear Research, 141980~Dubna, Russia}\\
\mbox{$^{7}$Institut f\"{u}r Kernphysik, Johann Wolfgang Goethe-Universit\"{a}t,
60438 ~Frankfurt, Germany}\\
\mbox{$^{8}$II.Physikalisches Institut, Justus Liebig Universit\"{a}t Giessen,
35392~Giessen, Germany}\\
\mbox{$^{9}$Istituto Nazionale di Fisica Nucleare, Sezione di Milano,
20133~Milano, Italy}\\
\mbox{$^{10}$Institute for Nuclear Research, Russian Academy of Science,
117312~Moscow, Russia}\\
\mbox{$^{11}$Physik Department E12, Technische Universit\"{a}t M\"{u}nchen,
85748~M\"{u}nchen, Germany}\\
\mbox{$^{12}$Department of Physics, University of Cyprus, 1678~Nicosia, Cyprus}\\
\mbox{$^{13}$Institut de Physique Nucl\'{e}aire (UMR 8608),
CNRS/IN2P3 - Universit\'{e} Paris Sud, F-91406~Orsay Cedex, France}\\
\mbox{$^{14}$Nuclear Physics Institute, Academy of Sciences of Czech Republic,
25068~Rez, Czech Republic}\\
\mbox{$^{15}$Departamento de F\'{\i}sica de Part\'{\i}culas, Univ. de Santiago de Compostela,
15706~Santiago de Compostela, Spain}\\
\mbox{$^{16}$Instituto de F\'{\i}sica Corpuscular, Universidad de Valencia-CSIC,
46971~Valencia, Spain}\\
\mbox{$^{a}$ also at ISEC Coimbra, Coimbra, Portugal}\\
\mbox{$^{b}$ also at Technische Universit\"{a}t Dresden,
01062~Dresden, Germany}\\
\mbox{$^{c}$ also at Excellence Cluster Universe, Technische Universit\"{a}t M\"{u}nchen,
85748 Garching, Germany}\\
}
\date{\today}

\begin{abstract}
Results on $\Lambda$p femtoscopy are reported at the lowest energy so far. At a beam energy of 1.76$A$~GeV, the reaction Ar+KCl was studied with the High Acceptance Di-Electron Spectrometer (HADES) at SIS18/GSI. A high-statistics and high-purity $\Lambda$ sample was collected, allowing for the investigation of $\Lambda$p correlations at small relative momenta. The experimental correlation function is compared to corresponding model calculations allowing the determination of the space-time extent of the $\Lambda$p emission source. The $\Lambda$p source radius is found slightly smaller than the pp correlation radius for a similar collision system. The present $\Lambda$p radius is significantly smaller than that found for Au+Au/Pb+Pb collisions in the AGS, SPS and RHIC energy domains, but larger than that observed for electroproduction from He. Taking into account all available data, we find the $\Lambda$p source radius to increase almost linearly with the number of participants to the power of one-third.

\end{abstract}


\maketitle
Two-particle correlations of hadrons at small relative momenta are widely used to study the space-time extent of their source created in heavy-ion collisions or other reactions involving hadrons (for review see \cite{Lisa05}). Generally, the sign and strength of the correlation is affected by i) the strong interaction, ii) the Coulomb interaction if charged particles are involved, and iii) the quantum statistics in case of identical particles (Pauli suppression for fermions, Bose-Einstein enhancement for bosons). Their interplay makes, e.g., pp correlation functions rather complex with a suppression at very small relative momenta (due to items ii) and iii)) followed by a maximum which results from the short-range attractive potential of their strong interaction. In contrast, correlations of nonidentical hadrons, with at least one partner being uncharged, are sensitive to the strong interaction alone. Hence, $\Lambda$p correlation functions are well suited to study the spatiotemporal extension of the particle emitting source, provided the $\Lambda$p interaction is known \cite{WangPratt99}. Presently, information on $\Lambda$p correlations is rather scarce. Due to the necessity of having high statistics of protons and $\Lambda$'s at small relative momenta, such information is available at high beam energies only. Experimental $\Lambda$p correlation functions have been reported for central collisions of Au+Au/Pb+Pb by E895 \cite{AGS03_E895} at the Alternating Gradient Synchrotron (AGS) at Brookhaven National Laboratory (BNL), by NA49 \cite{SPS03_NA49} at the Super Proton Synchrotron (SPS) at CERN, by STAR \cite{RHIC05_STAR} at the Relativistic Heavy Ion Collider (RHIC) at BNL, and, for electroproduction from He, by CLAS \cite{CLAS09} at the Thomas Jefferson National Accelerator Facility (JLab). On the other hand, the higher the energy the wider the particles are distributed in momentum space resulting in a reduced probability of finding pairs with small relative momenta. Thus, at lower beam energies an exceptionally large number of events and the more compact momentum distribution of $\Lambda$ hyperons might compensate partially their lower production probability. Indeed, such a high-statistics $\Lambda$ sample is available to us \cite{Fabbietti_SQM08,HADES-Xi} making an analysis of $\Lambda$p correlations feasible.
Furthermore, at low energies the feed-down correction to the $\Lambda$ is smaller and more reliable, being of importance for comparisons to models applying explicitly to primary $\Lambda$'s like the model by Lednicky and Lyuboshitz \cite{Lednicky82} which we use in the present investigation.

In this paper we report on the first observation of $\Lambda$p correlations in heavy-ion collisions at beam energies below 2\,GeV per nucleon. The experiment was performed with the {\bf H}igh {\bf A}cceptance {\bf D}i-{\bf E}lectron {\bf S}pectrometer (HADES) \cite{hadesSpectro} at the Schwerionensynchrotron SIS18 at GSI, Darmstadt. HADES, primarily designed to measure di-electrons \cite{HADES-PRL07}, offers also excellent hadron identification capabilities \cite{PhD_Schmah,Fabbietti_SQM08,hades_kpm_phi_09,HADES-Xi}.

A $^{40}$Ar beam of about $10^6$ particles/s with kinetic energy of 1.756$A$~GeV ($\sqrt{s_{NN}}=2.61$~GeV) was incident on a four-fold segmented target of natural KCl with a total thickness of 5\,mm corresponding to $3.3\,\%$ interaction probability.
The data readout was started by a first-level trigger (LVL1) decision, requiring a minimum charged-particle multiplicity $\ge 16$ in the time-of-flight detectors covering polar angles from 18 to 85 degrees. The integrated cross section selected by this trigger comprises approximately the most central 35\,\% of the total reaction cross section. The corresponding mean number of participants, estimated with the UrQMD transport approach \cite{UrQMD} amounts to $\langle A_{part} \rangle=38.5 \pm 2.5$. About 700 million LVL1 events were processed for the present investigation.

Let $Y_{12}({\bf p}_1, {\bf p}_2)$ be the coincidence yield of pairs of particles having the momenta ${\bf p}_1$ and ${\bf p}_2$. Then the two-particle correlation function is defined as
\begin{equation}
\mbox{C}({\bf p}_1, {\bf p}_2) = {\cal N} \,
\frac{\sum _{events,pairs} Y_{12}({\bf p}_1, {\bf p}_2)}
{\sum_{events,pairs} Y_{12,mix}({\bf p}_1, {\bf p}_2)}.
\label{def_exp_corr_fct}
\end{equation}
The sum runs over all events and over all pairs satisfying certain conditions. Event mixing, denoted by the subscript ''mix'', means to take particle 1 and particle 2 from different events belonging to the same multiplicity interval of 8 units width \cite{PhD_Schmah}. Due to limited statistics, the six-dimensional correlation function is projected onto one-half of the relative momentum in the pair c.m. frame, i.e. we consider $\mbox{C}(k)$ with $k = \vert {\bf p}_1 - {\bf p}_2 \vert/2$. The normalization factor ${\cal N}$ in (\ref{def_exp_corr_fct}) is fixed by the requirement $\mbox{C}(k) \rightarrow 1$ at large relative momenta, $k= 0.10$ - 0.15\,GeV/c. The statistical errors of $\mbox{C}(k)$ are governed by those of the coincidence yield, since the mixed yield can be generated with much higher statistics.

\begin{figure}[!htb]
\begin{center}
\includegraphics[width=1.\linewidth,viewport=0 0 570 530]{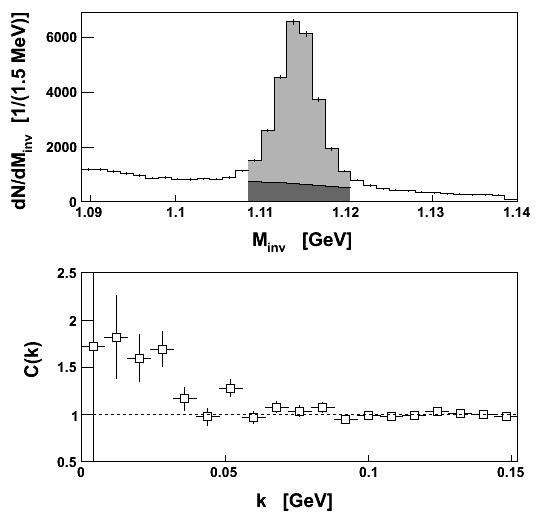}
\caption[]{Top: The p$\pi^-$ invariant mass distribution. Light and dark hatched areas represent the statistics of the $\Lambda$ signal and the corresponding $\Lambda$ impurity, respectively, both contributing to the $\Lambda$p correlation analysis. Bottom: The uncorrected $\Lambda$p correlation function.
\label{LambdaMass_pLamRaw}}
\end{center}
\end{figure}
In the present analysis we identified the $\Lambda$ hyperons through their decay $\Lambda \rightarrow \mathrm{p} \pi^-$. To allow for a clear $\Lambda$ selection various cuts were applied (cf. \cite{HADES-Xi}). The resulting invariant mass distribution of events containing at least one additional well-identified proton is displayed in the upper panel of Fig.\,\ref{LambdaMass_pLamRaw}.

Taking the resulting $\Lambda$ sample (hatched area, $\pm 6$\,MeV interval around $\Lambda$ peak), we started the correlation study by combining, for each event containing a $\Lambda$ candidate, the $\Lambda$ with those protons not already contributing to the $\Lambda$. The result is a $\Lambda$p relative-momentum distribution containing a total of 240,000 proton-$\Lambda$ pairs. However, only 2,700 (260) pairs contribute to the interesting region of small relative momenta, $k<0.1\,(0.04)$\,GeV/c. The resulting raw $\Lambda$p correlation function is displayed in the lower panel of Fig.\,\ref{LambdaMass_pLamRaw}. A clear enhancement at small relative momenta is found, indicating the effect of the strong $\Lambda$p interaction.

\begin{figure}[!htb]
\begin{center}
\includegraphics[width=1.\linewidth,viewport=0 0 570 530]{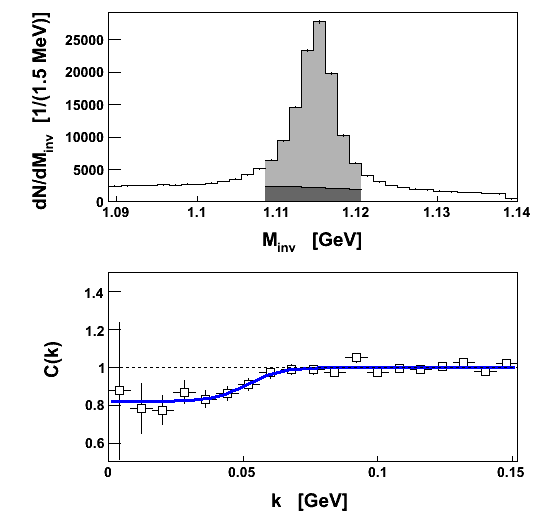}
\caption[]{The same as Fig.\,\ref{LambdaMass_pLamRaw}, but for simulated $\Lambda$ hyperons with their decay products embedded into true experimental data. The full curve in the lower panel is a fit with a Fermi-type function (see text) used to correct for reconstruction losses (due to close tracks) of the experimental $\Lambda$p correlation function displayed in Fig.\,\ref{LambdaMass_pLamRaw}.
\label{pLamEmb} }
\end{center}
\end{figure}
Corrections for the finite small-angle acceptance losses were deduced from simulations. 25 million thermal $\Lambda$'s (one per event) were generated using the fireball option of the event generator PLUTO \cite{Pluto}. The fireball parameters of the $\Lambda$ are chosen such that the simulation reproduces both the experimental values of the effective inverse slope parameter at mid-rapidity and the rapidity width \cite{Fabbietti_SQM08,HADES-Xi,PhD_Schmah}. The simulation data are processed through GEANT \cite{GEANT}, modeling the detector response. The GEANT data were embedded into real experimental data and processed through the full analysis chain. The correlation function of, ab initio, uncorrelated pairs of (simulated) $\Lambda$ hyperons and (experimental) protons is expected to be flat at unity. Any deviation from this value has to be interpreted as detector bias due to finite granularity, momentum and near-track resolutions. Figure\,\ref{pLamEmb} shows the correlation function of simulated $\Lambda$'s combined with protons from experimental data. A slight suppression at small relative momenta is visible indicating the bias of the apparatus and the track-finding losses. The full curve is a fit with a Fermi-type function, $f_{acc}(k)= 1 + A_1/(1+e^{(k-A_2)/A_3})$, used to correct the experimental $\Lambda$p correlation function for reconstruction losses due to close tracks. The close-track correction is performed via $\mbox{C}(k) \rightarrow \mbox{C}(k)/f_{acc}(k)$, while the purity correction is done by the transformation $\mbox{C}(k) \rightarrow 1 + (\mbox{C}(k)-1)/PairPurity$. Here, $PairPurity$ is the product of the proton ($0.95\pm0.02$) \cite{PhD_Schmah} and $\Lambda$ purities.
The $\Lambda$ purity, in turn, follows from the product of the signal purity of $0.816 \pm 0.013$ as derived from the p$\pi^-$ invariant mass distribution of Fig.\,\ref{LambdaMass_pLamRaw} (which increases to $0.876\pm0.014$ for half the given mass window) and the feed-down correction of the decay $\Sigma^0\rightarrow \Lambda \gamma$. (Feed down from decays of heavier particles is neglected in the present investigation.) On the one hand, the shape and the magnitude of the $\Sigma^0$p and $\Lambda$p correlation functions are predicted to be close to each other \cite{Mikhaylov06}. On the other hand, residual correlations carried by secondary $\Lambda$'s are expected to be partially dissolved by decay kinematics. Hence, we approximated the feed-down correction midway between these boundaries, yielding a factor of $0.90 \pm 0.10$. Here, the upper limit of the correction factor (i.e. no correction) is related to identical $\Lambda$p and $\Sigma^0$p correlations combining into the measurable one, while the lower limit meets the case of maximum feed-down correction, i.e. assuming a complete loss of the $\Sigma^0$p correlation after the electromagnetic decay of the $\Sigma^0$. The corresponding yield ratio of $\Lambda/(\Lambda + \Sigma^0)$ of about 0.80 was derived from calculations with the UrQMD transport approach \cite{UrQMD}. At last, utilizing GEANT simulations, the influence of the momentum resolution onto the $\Lambda$p correlation function was investigated.  It appears to be negligible (i.e. less than 2\,\% increase of the width of $\mbox{C(k)}-1$, which can be approximated by a Gaussian function with a dispersion of $(27\pm3)$\,MeV/c), as expected from the narrow widths of both the $\Lambda$ and $\Xi^-$ signal peaks \cite{HADES-Xi}.

\begin{figure}[!htb]
\begin{center}
\includegraphics[width=0.52\linewidth,viewport=0 0 530 530]{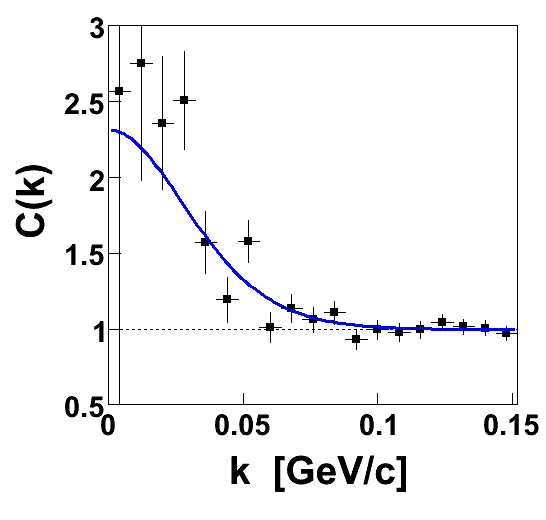} \hfill
\includegraphics[width=0.47\linewidth,viewport=0 0 530 530]{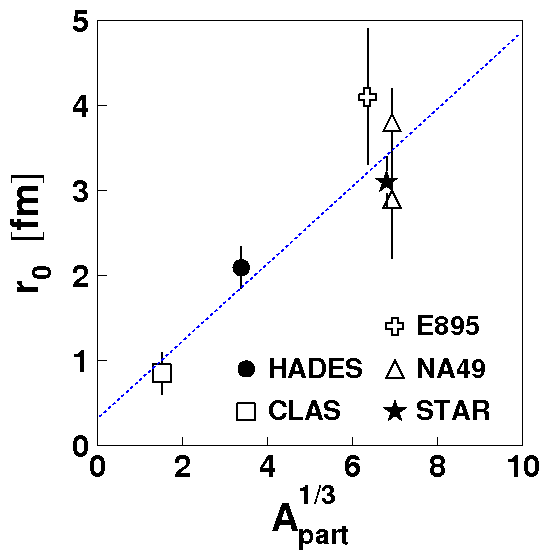}
\caption[]{Left: The $\Lambda$p correlation function after close-track and purity corrections. The full curve represents the best fit (see text) with the Analytical Model by Lednicky and Lyuboshitz \cite{Lednicky82}. Right: The Gaussian radius of the $\Lambda$p emission source as a function of system size. The symbols indicate data taken with HADES at SIS, CLAS \cite{CLAS09} at JLab, E895 \cite{AGS03_E895} at AGS, NA49 \cite{SPS03_NA49} at SPS, and STAR \cite{RHIC05_STAR} at RHIC, respectively. The dashed line is a linear regression to the data.
\label{pLamFitLednicky} }
\end{center}
\end{figure}

Finally, we applied a fit to the corrected $\Lambda$p correlation function (Fig.\,\ref{pLamFitLednicky}, left panel) with a fit function derived from \cite{Lednicky82}. Following the procedure described in \cite{RHIC05_STAR}, we used the $\Lambda$p scattering lengths ($f_0^s=-2.88$\,fm, $f_0^t=-1.66$\,fm) and effective ranges ($d_0^s=2.92$\,fm, $d_0^t =3.78$\,fm) for the spin singlet and triplet states of the $\Lambda$p system as given in \cite{WangPratt99}. Notice that by increasing the source radius both the height and the width of the correlation above unity decrease. We applied the new log-likelihood minimization procedure recently proposed in \cite{Ahle02} the result of which, however, deviates hardly from that of a $\chi^2$ minimization. The optimum Gaussian radius provided by this fit amounts to
\begin{equation}
 r_0=(2.09 \pm 0.16\,^{+0.12}_{-0.10}\,^{+0.09}_{-0.16}\,^{+0.09}_{-0.11})\,\mbox{fm},
\label{R0result}
\end{equation}
where the 1st error is the statistical error, while the 2nd, 3rd, and 4th ones represent the systematic errors due to the uncertainties of the close-track correction with embedded $\Lambda$'s, due to the pair purity correction (with a systematically smaller radius by 0.12\,fm for half the $\Lambda$ mass window as given in Fig.\,\ref{LambdaMass_pLamRaw}), and due to a $\pm 25$\,\% variation of the scattering lengths entering the model \cite{Lednicky82}, respectively.
(Note that a fit to the uncorrected correlation function as displayed in the lower panel of Fig.\,\ref{LambdaMass_pLamRaw} would give a Gaussian source radius of 3.1\,fm.)

We studied the source radius resulting from the fit when changing the whole set of $\Lambda$p singlet and triplet scattering lengths and effective ranges describing the $\Lambda$p potential. In ref.\,\cite{Haidenbauer07} a compilation of such parameters is given. For the sets by the chiral effective field theory EFT\,'06, the J\"ulich\,'04 and Nijmegen NSC97f models the fit yields a radius of $(2.13\pm0.15)$\,fm, $(2.15\pm0.15)$\,fm and $(2.11\pm0.15)$\,fm, respectively. Though there are quite substantial differences in the parameter sets, the fitted source radii do not show serious differences and are very similar to the value (\ref{R0result}) derived above for the set from \cite{WangPratt99}. Obviously, any sensitivity to the peculiarities of the $\Lambda$p two-body interaction is dissolved after the integration over the corresponding wave functions and the source distribution, i.e. the procedure yielding the theoretical correlation functions. Hence, the correlation observed in the $\Lambda$p momentum difference is primarily sensitive to the source size.

The $\Lambda$p source radius (\ref{R0result}) appears slightly smaller than the pp source radii derived by the FOPI collaboration \cite{FOPI05} comparing two-proton small-angle correlations in central collisions of Ca+Ca to the predictions of the Koonin model \cite{Koonin77}. 

The present $\Lambda$p source radius may be compared to the corresponding radii derived in other experiments. In the right panel of Fig.\,\ref{pLamFitLednicky} we show the Gaussian radius as a function of the number of participants to the power of one-third, $A_{part}^{1/3}$, which is calculated from the centrality and the total size of the corresponding collision system using a geometrical model of penetrating sharp spheres. While for the data measured by NA49 \cite{SPS03_NA49} at SPS (158$A$\,GeV Pb+Pb, preliminary), by STAR \cite{RHIC05_STAR} at RHIC (Au+Au at $\sqrt{s_{NN}}=200$\,GeV), and by CLAS \cite{CLAS09} at JLab (preliminary results from $e$+$^3$He\,($^4$He) at 4.7\,(4.46)\,GeV),  the Gaussian radius $r_0$ is determined using the same model as in the present analysis, the half-maximum radius $R_{1/2}$ derived by E895 \cite{AGS03_E895} at AGS (6$A$\,GeV Au+Au) applying an imaging procedure was transformed to a Gaussian radius via $r_0=R_{1/2}/\sqrt{2\ln{2}}$. Clearly, the $\Lambda$p source radius increases with system size. Similarly to the systematics of two-pion \cite{Lisa05,PHENIX_pipi_04} and two-kaon \cite{PHENIX_kk_09} source radii, we find an almost linear increase with $A_{part}^{1/3}$. Finally, we point out that the precision, i.e. the relative error, of our $\Lambda$p source radius (\ref{R0result}) can well compete with or is even better than the accuracies reported at higher energies. Nevertheless, with presently available statistics it is not possible to use the data sample for getting deeper insight into the specifics of the $\Lambda$p interaction.

In summary, we observed, for the first time in heavy-ion collisions at SIS energies, $\Lambda$p correlations at small relative momenta. The system Ar+KCl at 1.76$A$~GeV beam energy was measured with the HADES detector. The $\Lambda$p correlation function, after proper corrections, was compared to the output of the model by Lednicky and Lyuboshitz. The resulting Gaussian source radius is slightly smaller than that derived from pp correlations of a similar collision system, when interpreted with the Koonin model. The present $\Lambda$p source radius is found significantly smaller than that deduced from Au+Au/Pb+Pb collisions at higher beam energies and larger than that reported for electroproduction from He. It increases almost with $A_{part}^{1/3}$. A significant variation of the scattering lengths and the effective ranges of the $\Lambda$p interaction entering the model changes only marginally the source radius resulting from the one-parameter fit to the experimental $\Lambda$p correlation function. This indicates the robustness of the correlation function w.r.t. uncertainties in the knowledge of the $\Lambda$p two-body interaction. Finally, we note that the envisaged high-intensity runs of both the upgraded HADES and the Compressed Baryon Matter Experiment planned at the future Facility for Antiproton and Ion Research (FAIR) \cite{FAIR} may provide the statistics necessary for more differential correlation analyses even of pairs of rare particle species, hence gaining information not only on the extents of the emission sources but also on their
shape or spatiotemporal separation.

\vspace*{10bp}
The HADES collaboration gratefully
acknowledges the support by BMBF grants 06DR9059D, 06FY171, 06MT238~T5, and 06MT9156~TP5,
by HGF VH-NG-330, by DFG EClust 153, by GSI TMKRUE,
by the Hessian LOEWE initiative through HIC for FAIR (Germany),
by grants MSMT LC07050 and GA ASCR IAA100480803 (Czech Rep.),
by grant KBN 1P03B 056 26 (Poland),
by grants FPA2006-09154 and CPAN:CSD2007-00042 (Spain),
by grant UCY-10.3.11.12 (Cyprus), by CNRS/IN2P3 (France), by INFN (Italy),
and by EU contract RII3-CT-2005-515876.


\end{document}